\documentclass[12pt]{article}
\usepackage{epsfig}
\usepackage{amsfonts}
\usepackage{amssymb}

\newcommand{\eqn}[1]{(\ref{#1})}
\newcommand{\be}{\begin{equation}}
\newcommand{\ee}{\end{equation}}
\newcommand{\ben}{\begin{displaymath}}
\newcommand{\een}{\end{displaymath}}
\newcommand{\bea}{\begin{eqnarray}}
\newcommand{\eea}{\end{eqnarray}}
\newcommand{\bean}{\begin{eqnarray*}}
\newcommand{\eean}{\end{eqnarray*}}
\newcommand{\nn}{\nonumber \\}
\newcommand{\ba}{\begin{array}}
\newcommand{\ea}{\end{array}}
\newcommand{\bi}{\begin{itemize}}
\newcommand{\ei}{\end{itemize}}

\newcommand{\reef}[1]{(\ref{#1})}
\newcommand{\non}{\nonumber}

\def\G{\Gamma}

\def\e{\epsilon}

\renewcommand{\t}{\theta}

\def\bnum{\begin{enumerate}}
\def\enum{\end{enumerate}}

\def\CR{\mathbb{R}}
\def\CM{\mathcal{M}}

\def\8M{$\CM_8$}

\def\be{\begin{equation}}
\def\ee{\end{equation}}
\def\G{\Gamma}

\def\ei{e^{\underline{i}}}

\def\e1{e^{\underline{1}}}
\def\1u{\underline{1}}
\def\2u{\underline{2}}

\def\0u{\underline{0}}
\def\e{\epsilon}
\def\target{$\CR^{1,1}\times \mathcal{M}_8$ }
\def\target2{$\CR^{1,1}\times \mathcal{M}_8$,}
\def\9G{\G_{\underline{9}}}

\def\p{\partial}

\newcommand{\cala}{\mbox{${\cal A}$}}
\newcommand{\calb}{\mbox{${\cal B}$}}

\newcommand{\calf}{\mbox{${\cal F}$}}

\newcommand{\calj}{\mbox{${\cal J}$}}

\newcommand{\bbr}[1]{\mbox{${\mathbb R}^{#1}$}}

\newcommand{\qt}{\tilde{q}}

\newcommand{\rom}[1]{\mathrm{#1}}
\newcommand{\ti}[1]{\tilde{#1}}
\newcommand{\mc}[1]{\mathcal{#1}}

\newcommand{\pa}{\partial}

\newcommand{\sac}{\, , \qquad}
\newcommand{\eg}{{\it e.g. }}
\newcommand{\ie}{{\it i.e. }}

\newcommand{\ra}{\rightarrow}

\begin{document}

\begin{titlepage}
\begin{flushright}
hep-th/0504125
\end{flushright}

\vskip 1.5in
\begin{center}
{\bf\large {Supersymmetric 4D Rotating Black Holes from 5D Black
Rings}}
\vskip 0.5in
{Henriette Elvang$^{a}$, Roberto Emparan$^{bc}$, David Mateos$^{d}$
and Harvey S.~Reall$^{e}$}
\vskip 0.3in
{\small {\textit{$^{a}$Department of Physics, University of
California, Santa Barbara, CA 93106-9530, USA}}}
\\ {\small{
\textit{ $^b$Instituci\'o Catalana de Recerca i Estudis Avan\c{c}ats (ICREA)\\
$^c$Departament de F{\'\i}sica Fonamental, and \\
C.E.R. en Astrof\'{\i}sica, F\'{\i}sica de Part\'{\i}cules i
Cosmologia,
Universitat de Barcelona, Diagonal 647, E-08028 Barcelona, Spain}}}
\\ {\small{
\textit{ $^d$Perimeter Institute for Theoretical Physics, Waterloo, Ontario N2L 2Y5, Canada}}}
\\ {\small{
\textit{ $^e$Kavli Institute for Theoretical Physics, University of
California, Santa Barbara, CA 93106-4030, USA}}}

\vspace*{0.2cm}
{\tt elvang@physics.ucsb.edu, emparan@ub.edu,}\\
{\tt dmateos@perimeterinstitute.ca, reall@kitp.ucsb.edu}\\[5mm]
NSF-KITP-05-25
\end{center}
\vspace*{0.2cm}

\baselineskip 16pt
\date{}

\begin{abstract}
We present supersymmetric solutions describing black holes with
non-vanishing angular momentum in four dimensional asymptotically flat
space. The solutions are obtained by Kaluza-Klein reduction of
five-dimensional supersymmetric black rings wrapped on the fiber of a
Taub-NUT space. We show that in the four-dimensional description the
singularity of the nut can be hidden behind a regular black hole event
horizon and thereby obtain an explicit example of a non-static multi-black
hole solution in asymptotically flat four dimensions.
\end{abstract}
\vfill \setcounter{page}{0} \setcounter{footnote}{0}
\end{titlepage}
\vfill\eject

\tableofcontents

\setcounter{equation}{0}
\section{Introduction}

It is well-known that when the supersymmetric limit of a
four-dimensional Kerr-Newman black hole is taken, one must at the
same time set the angular momentum to zero in order to preserve the
regular horizon.\footnote{See, \eg \cite{PKT} for a general
discussion on the interplay between angular momentum and
supersymmetry. Our work indicates that still more surprises with
angular momentum remain to be discovered.} Thus, no rotating,
supersymmetric, single-black hole solutions have been known in
asymptotically flat four-dimensional spacetimes.

On the other hand, supersymmetric rotating black holes {\it do}
exist in five dimensions. Besides the BMPV black hole of
\cite{BMPV}, with a horizon of spherical topology, recently we
constructed a supersymmetric black ring of five-dimensional
supergravity \cite{EEMR1}. In this paper, we shall show that there
is a natural mechanism that reduces these five-dimensional black
rings to supersymmetric configurations of four-dimensional black
holes with angular momentum. The existence of the configuration that
we shall describe has been conjectured by the authors of \cite{BK3}.
A direct construction in four dimensions of supersymmetric
multi-black hole solutions with angular momentum has been given
earlier in \cite{denef}.\footnote{See \cite{BLS} for an early
general study of stationary supersymmetric solutions in four
dimensions.}

In a four-dimensional set-up, the basic idea is as follows. Consider
a supersymmetric, asymptotically flat black hole solution of
${\cal N} =2$ supergravity which is electrically
charged with respect to one of the gauge fields of the theory. If we
now place a magnetic monopole, of the same (Abelian) gauge field, a
finite distance away from the black hole, then angular momentum is
generated by the crossed electric and magnetic fields. Properly
speaking, the black hole itself is not rotating, since the geometry
near the horizon is static,\footnote{Note that a supersymmetric,
asymptotically flat, black hole must have vanishing angular velocity
but nevertheless one often refers to such an object as rotating if
it is non-static. Examples are the 5D BMPV black hole \cite{BMPV,
GMT} or the supersymmetric black ring
\cite{EEMR1}.} but the configuration is able to carry a non-vanishing
angular momentum while preserving four supercharges. It therefore seems
appropriate to refer to the system as rotating.

In the simplest form of this configuration, the magnetic monopole
gives rise to a naked singularity. So one may worry that such
singularities are the price to pay in order to have both rotation
and supersymmetry. However, we shall show that our five-dimensional
construction can be easily extended to hide away the singularity
behind a black hole horizon, and construct a supersymmetric
two-black hole configuration with non-zero angular momentum. This is
free from naked singularities as well as from other possible
pathologies such as closed timelike curves (CTCs).\footnote{Such
regular, asymptotically flat, non-static multi-black hole solutions
apparently do not exist in {\it minimal}
${\cal N}=2$ supergravity (\ie Einstein-Maxwell theory). Assuming that
any such solution must be supersymmetric, the results of \cite{tod}
reveal that it would have to belong to the Israel-Wilson-Perjes family
of solutions but it has been argued that the only black hole solutions
of this family are the static multi-Reissner-Nordstrom solutions
\cite{hartle}.}

We obtain the four-dimensional solutions by Kaluza-Klein (KK) reduction
of five-dimen\-sional solutions consisting of supersymmetric black rings
\cite{EEMR1}
in a Taub-NUT (TN) space. A rough way to think about this configuration
(which is explained more precisely in section \ref{s:4dred}) is that the
$S^1$ factor of the black ring horizon wraps the $S^1$ fiber of the TN
space, so reduction to four dimensions yields a black hole with an
$S^2$ horizon. The black ring carries two independent angular
momenta. One combination of these descends to the angular momentum
of the four-dimensional black hole, whereas the other corresponds to
momentum along the TN fiber and so becomes the electric charge of
the four-dimensional black hole. Although the five-dimensional
solution is completely regular (up to a harmless orbifold
singularity), the `nut' of the TN space yields a naked singularity
of a KK monopole upon reduction to four dimensions. This can be
remedied by placing at the nut a five-dimensional black hole with an
$S^3$ horizon, which is mutually supersymmetric with the black ring.
Reduction to four dimensions then yields a solution with two regular
$S^2$ horizons and non-zero angular momentum.

A closely related configuration has been considered in \cite{BK3}
where a two-charge supertube \cite{mt}, instead of a black ring,
wraps the TN fiber. Our configuration can be regarded as an
extension of this to three-charge/three-dipole supertubes
\cite{EEMR2,BW,GG2}, which are known to correspond, for equal
charges, to the supersymmetric black rings that we consider in this
paper. In fact, it should be straightforward to extend our solution,
using the methods in \cite{GG2}, to general three-charge supertubes
in TN. However, only in the case that the supertube carries all
three charges and three dipoles does one obtain a solution with a
regular horizon, both in 4D and 5D.

The ability of Taub-NUT space to compactify a five-dimensional black
object to a four-dimensional one has also been exploited very recently
in \cite{GSY}. In this case a black hole (essentially a Taub-NUT
extension of the BMPV solution) is placed at the nut with the rotation
one-form aligned with the TN fiber, so the resulting four-dimensional
solution does not carry any angular momentum. One limit of our
solutions, obtained by collapsing the ring to the nut, also describes a
black hole in Taub-NUT. However, as we shall discuss in section
\ref{s:4dsoln}, it is different than the one in \cite{GSY}.

The rest of the paper is organized as follows. In section
\ref{s:soln} we construct the five-dimensional solution describing a
supersymmetric black ring in a Taub-NUT background. We study its
physical properties in section \ref{s:physprop}: this includes the
near-horizon geometry (section \ref{s:nh}), the non-trivial
asymptotic structure (section \ref{s:5dasymp}), the conditions for
absence of CTCs (section \ref{s:noCTCs}), and the physical
parameters (section \ref{s:5dcharges}). In section
\ref{s:limits} we study briefly two particular limits of the system: the
collapse of the ring to a black hole at the nut, and the removal of the
nut to infinity, in which the ring becomes a compactified black string. We
perform the reduction to four dimensions in section \ref{s:4dred} and
study the properties of the four-dimensional black hole obtained from
the five-dimensional black rings. In section \ref{s:coverNUT} we
construct a five-dimensional solution describing, in the Taub-NUT
background, a black hole at the nut and a black ring away from the nut.
We conclude with a discussion in section \ref{s:discussion}.

The appendices contain technical details. Appendix \ref{app:xy} provides
the black ring solution in coordinates useful for studying the
near-horizon structure, and for connecting the solutions found here with
the supersymmetric black rings in asymptotically flat space. In appendix
\ref{app:horizon} we prove that the black ring horizon is smooth, and in
appendix \ref{app:noCTCs} we show that provided the parameters obey
certain constraints there are no CTCs outside the horizon.

\medskip

{\em Note added:} Following the appearance of the first version of this
paper, we were informed by F.~Denef of his earlier four-dimensional
construction of supersymmetric rotating multi-black holes \cite{denef},
of which we were unaware. Shortly afterwards ref.~\cite{GSY2} appeared,
which also describes black rings in Taub-NUT and provides the connection
between the two constructions. More recently, TN-black rings have also
been discussed in \cite{BKW}.

\setcounter{equation}{0}
\section{Black Rings in Taub-NUT space} \label{s:soln}

Any supersymmetric solution of five-dimensional minimal supergravity
must admit a non-spacelike Killing vector field $V$ \cite{gibbons:94}. In a region
where $V$ is time-like, the metric and gauge potential can be
written as
\cite{harveyetal}
\bea
 ds^2 &=& - H^{-2} \, (dt+\omega)^2 + H \, ds^2(\calb) \,,
\label{metric} \\
 \mc{A}&=&\frac{\sqrt{3}}{2}\left[H^{-1}(dt+\omega)-\beta\right]\,,
\label{potential}
\eea
where $V = \partial/\partial t$, the so-called base space $\calb$ is
an arbitrary hyper-K\"ahler space, and $H$ is a scalar function while $\omega, \beta$
are one-forms on $\calb$. The one-form $\beta$ is related to $\omega$ through the
relation $3 H d\beta= d\omega + \star_4 d\omega$, where $\star_4$ is the Hodge
star on $\calb$.

Supersymmetry implies that $H$ and $\omega$ must obey a pair of coupled
equations on $\calb$. If $\calb$ is a Gibbons-Hawking (GH) space
\cite{GH}, i.e, if $\calb$ admits a Killing field that preserves the
complex structures, then it is straightforward to solve these
equations.\footnote{Provided one assumes that the GH Killing field
extends to a symmetry of the full 5D solution.} The general solution
involves four harmonic functions on $\bbr{3}$ \cite{harveyetal}.

Recently, a supersymmetric, asymptotically flat, black ring solution was
obtained \cite{EEMR1}. The base space of this solution is flat:
$\calb=\bbr{4}$. It was observed in \cite{GG1} that this solution can be
written in the GH form just discussed. Now, (self-dual, Euclidean)
Taub-NUT space is also a GH space. This suggests that we should be able
to construct a solution describing a supersymmetric black ring in
Taub-NUT as a special case of the general GH solution of
\cite{harveyetal}.

The metric of Taub-NUT space is
\bea
  \label{TNmetric}
  ds^2(\rom{TN}) &=&
  H_k \, [dr^2 + r^2 d\theta^2 + r^2 \sin^2 \theta \, d\phi^2] +
  H_k^{-1} \, (dz+ Q_k \cos \theta \, d\phi)^2 \,,
\eea
with
\be
H_k = 1+\frac{Q_k}{r}
\ee
and $0\le \theta \le \pi$. The periodicities of the angular
coordinates are $\Delta\phi=2\pi$ and $\Delta z=2\pi R_k$. The parameter
$Q_k$ is quantized as
\be
Q_k=\frac{1}{2} N_k R_k\,,
\ee
where $N_k$ is an integer. Note that $R_k$ is the
asymptotic radius of the $z$-circle in the base space, but it need not (indeed it will not)
be the asymptotic radius in the full solution. Similarly, $Q_k$ will not
be exactly the same as the magnetic KK charge. The surfaces of constant
$r$ are squashed lens spaces $S^3/\mathbb{Z}_{N_k}$.

The solution \reef{metric}-\reef{potential} is specified in terms of three harmonic
functions $K$, $L$ and $M$ on $\bbr{3}$ as follows (the fourth harmonic function is $H_k$).
First, the one-forms $\beta$ and $\omega$ are split as
\be\label{beom}
\beta = \beta_0 (dz + Q_k\cos\theta d\phi) + \tilde{\beta}
\sac \omega = \omega_0 (dz + Q_k\cos\theta d\phi) +
\tilde{\omega} \,,
\ee
where $\tilde{\beta}$ and $\tilde{\omega}$ are one-forms on
$\bbr{3}$. Then $H$, $\beta_0$ and $\omega_0$ are given by \cite{harveyetal,gmr}
\bea
H &=& H_k^{-1} K^2 + L \,, \nn
\beta_0 &=& H_k^{-1} K \,, \nn
\omega_0 &=& H_k^{-2} K^3 + \frac{3}{2} H_k^{-1} K L + M \,,
\label{LKM}
\eea
whereas $\tilde{\beta}$ and $\tilde{\omega}$ are determined by
\bea
d\ti{\beta} &=& - \star_3 dK \, ,\non \\
  d\ti{\omega} &=& \star_3
  \left[H_k dM - M d H_k + \frac{3}{2}(KdL - LdK)\right] \, ,
\label{omega}
\eea
where $\star_3$ denotes the Hodge dual on $\bbr{3}$.

Guided by \cite{GG1}, we obtain a supersymmetric black ring in Taub-NUT by choosing the harmonic functions to be
\be
K = - \frac{q}{2\Delta} \sac
L = 1 + \frac{Q- 2 q v_H Q_k}{4Q_k \Delta}\sac
M = v_H \left(1 - \frac{R}{\Delta}\right)
\,,
\label{tnKLM}
\ee
where $q$, $Q$ and $R$ are constants,
$\Delta$ is the distance to
$\vec{x}_0=(0,0,-R)$ in $\bbr{3}$,
\be\label{harmonic}
\Delta=|\vec{x} - \vec{x}_0| = \sqrt{r^2 + R^2 + 2 R r
\cos{\theta}} \,
\ee
and we have defined
\bea\label{vH}
  v_H=\frac{3q}{4 \, (Q_k+R)}\, .
\eea
The parameters $R$, $Q_k$, $q$ have dimensions of length and $Q$ has
dimensions of length squared. $Q$ and $R$ have the interpretation of the
charge of the ring and (a measure of) its distance to the nut, but it
should be noted that they do not bear a simple direct relation to the
parameters with the same labels in the solution for the ring in
asymptotically flat 5D space in \cite{EEMR1}. $q$ yields the dipole
charge of the ring and is essentially the same parameter as in
\cite{EEMR1}. The detailed relation between this solution and the one in
\cite{EEMR1} is given in appendix \ref{app:xy}. In the next section we
shall see that the dimensionless quantity $v_H$ can be interpreted as a
velocity of the five-dimensional horizon in the $z$ direction.

We shall assume that
\bea
  Q\ge  2 q v_H Q_k\, ,~~~~~
  q\ge 0 \, .
\eea
We can now collect these pieces to give the solution we seek. With
the above harmonics \reef{tnKLM} we have
\bea
  H = 1 +
\frac{Q- 2 q v_H Q_k}{4Q_k \Delta}
  + \frac{q^2}{4\, H_k\, \Delta^2} \, .
\eea
The one-forms $\omega$ and $\beta$ are found using \reef{LKM} and \reef{omega}. In
particular,
\be
\omega_0=v_H\left(1-\frac{R}{\Delta}\right)-
\frac{3q}{4H_k \Delta}\left(1 +
\frac{Q- 2 q v_H Q_k}{4Q_k \Delta}\right)
-\frac{q^3}{8H_k^2\Delta^3}
\ee
and
\be
  \ti{\omega}=\ti{\omega}_\phi \, d\phi= \frac{2v_H\, Q_k R\,  r\,  \sin^2\theta}
   {\Delta \, (r+R+\Delta)} \, d\phi\,.
\label{tilde}
\ee
$\beta$ is simply
\be\label{beta}
\beta=-\frac{q}{2\Delta}\left[H_k^{-1}(dz+Q_k \cos\theta \,d\phi)-
(R+r\cos\theta)d\phi\right]\,.
\ee
Together with \reef{TNmetric}, these provide all the input needed to specify
the solution in \reef{metric}, \reef{potential}.

We have made a number of choices of various coefficients in the harmonic
functions and of integration constants. The choices are made to ensure
that the four-dimensional metric obtained by reduction along $\pa_z$ is
asymptotically flat and that the five-dimensional solution possesses no
Dirac-Misner strings. These two requirements translate into the
conditions that $\ti\omega_\phi \to 0$ as $r \to \infty$ and $\ti
\omega_\phi (\theta=0, \pi)=0$, respectively. We have furthermore made a
simplifying choice by having as well $K\to 0$ as $r\to\infty$, but this
can be relaxed. We comment on the
possibilities of more general solutions in section \ref{s:general}.

\setcounter{equation}{0}
\section{Physical Properties} \label{s:physprop}

The solution constructed in the previous section describes a supersymmetric
black ring with $S^2\times S^1$ horizon in a Taub-NUT background. In this section
we examine its physical properties as a five-dimensional solution; its reduction
to four dimensions will be studied in section \ref{s:4dred}.
First in subsection \ref{s:nh} we briefly discuss
the near-horizon limit. The details are fairly technical and are given in
appendix \ref{app:horizon}, where
we find coordinates that extend the metric smoothly through the horizon.
In section \ref{s:5dasymp} we study the non-trivial asymptotic
structure. Next in section \ref{s:noCTCs} we
comment on how to avoid CTCs (again the details are deferred to appendix
\ref{app:noCTCs}). Finally we compute the physical charges in
section \ref{s:5dcharges}.

\subsection{Near-horizon geometry} \label{s:nh}

Near the horizon, the structure of the solution remains basically the
same as for the black ring in asymptotically flat 5D of \cite{EEMR1}.
In the near-horizon limit (defined in appendix \ref{app:horizon}) we find
\bea\label{nhmetric}
  ds_\rom{nh}^2 = 2d\ti{v} d\ti{r}
  + \frac{4\hat{L}\ti{r}}{q} d\ti{v} d\hat{\psi}'
  + \hat{L}^2 d\hat{\psi}'^2
  + \frac{q^2}{4} \left[ d\bar\theta^2
  + \sin^2\bar\theta d\chi^2\right] \, .
\eea
Here $\ti{v}$ and $\hat{\psi}'$ are time and angular coordinates regular on the horizon,
while $\bar\theta$ and $\chi$ parametrize a 2-sphere. We have also
introduced
\be \label{ELL}
  \hat{L}=\sqrt{3\left[\left(\frac{Q- 2 q v_H Q_k}{2q}\right)^2-
\frac{4Q_k^2 R}{R+Q_k}
\right]}\,,
\ee
corresponding to the quantity defined in \reef{ELL2}.
The near-horizon limit of the metric
describes the product of locally AdS$_3$ of radius $q$ with a two-sphere
of radius $q/2$. This is exactly the same as for the asymptotically flat
black rings of \cite{EEMR1} up to the new expression for
$\hat{L}$.

A necessary condition for absence of CTCs near the horizon is
obviously $\hat{L}^2>0$, \ie
\be
\left(\frac{Q- 2 q v_H Q_k}{2q}\right)^2>\frac{4Q_k^2
R}{R+Q_k}\,.
\label{Qcondition}
\ee
This is sufficient to eliminate CTCs to leading
order near the horizon. However, in contrast to the situation for asymptotically flat
rings \cite{EEMR1}, this condition alone will \textit{not} be sufficient to avoid
causal pathologies in the full solution.

In appendix \ref{app:horizon} we show that both the metric and the inverse
metric are analytic at $\ti{r}=0$ and we show that the $\ti{r}=0$
hypersurface is indeed a Killing event horizon for the black ring.
The metric on a spatial cross-section of the horizon is
\bea
  ds_\rom{horizon}^2 =
  \hat{L}^2 d\hat{\psi}'^2
  + \frac{q^2}{4} \left[ d\bar\theta^2
  + \sin^2\bar\theta d\chi^2\right] \, ,
\eea
so the horizon topology is
$S^2 \times S^1$. The area of the horizon is
\be \label{ringarea}
A_\rom{H}=2\pi^2 N_k^{-1}\hat{L} q^2 =\frac{\pi^2 q}{N_k}
\sqrt{3\left[\left(Q- 2 q v_H Q_k\right)^2-16\frac{q^2Q_k^2 R}{R+Q_k}\right]}\,.
\ee
In appendix \ref{app:horizon} we show that the
black ring horizon is generated by the orbits of the Killing vector
\be\label{horgen}
  \xi= \frac{\partial}{\partial t}\,
\ee
at constant values of $z$ and $\phi$ (or $\hat\psi$ and $\hat\phi$ in
\reef{hatpsiphi}) on $\ti{r}=0$. This implies that the horizon velocity
in either of the directions $z$ or $\phi$ vanishes. However we will see
in the next subsection that surfaces of constant $z$ are moving with
respect to asymptotic observers at rest, which is in contrast to the
case of black rings in asymptotically flat 5d space \cite{EEMR1} where
the horizon is static.

\subsection{Asymptotic structure} \label{s:5dasymp}
The asymptotic structure of the Taub-NUT black rings turns out to be quite
non-trivial.
When $r\to \infty$, the 5d metric asymptotes to
\bea
 ds^2_\rom{asymp} =
 -\Big(dt + v_H(dz + Q_k\cos\theta \, d\phi) \Big)^2
 + (dz + Q_k\cos\theta \, d\phi)^2
 + d\mathbf{x}_3^2 \, .
\eea
The cross-term $dt(dz+Q_k\cos\theta\, d\phi)$ comes
from the fact that $\omega_0\to v_H$ asymptotically.
It implies that the asymptotic frame is not at rest. In order to go
to a frame that is asymptotically at rest, write first the asymptotic
metric as
\be \label{asym}
  ds_\rom{asymp}^2 =
  \gamma^{-2}
  \left[ (dz+Q_k \cos\theta \, d\phi)-v_H \gamma^2\, dt\right]^2
  -\gamma^2 dt^2
  + dr^2 + r^2 d\theta^2 +r^2 \sin^2 \theta \, d\phi^2 \, ,
\ee
where
\bea\label{gamdef}
  \gamma = \frac{1}{\sqrt{1- v_H^2}}\,.
\eea
We shall see shortly that $\gamma$ can be interpreted as a
relativistic dilation due to the horizon moving with velocity $v_H$
with respect to asymptotic observers. We now go to a frame at rest
by doing a coordinate transformation. At the same time, we rescale
$t$ and $z$ to be, respectively, canonically normalized time and
compact Kaluza-Klein direction at infinity:
\be
  t = \gamma^{-1}\, \bar{t} \, ,~~~~~
  z=\gamma \, (\bar{z}+v_H \, \bar{t}) \, .
  \label{asymptcoord}
\ee
The asymptotic metric is then
\bea \label{asym2}
  ds_\rom{asymp}^2 =
  \left[ d\bar{z}+\frac{Q_k}{\gamma}\cos\theta \, d\phi\right]^2
  - d\bar{t}^2
  +dr^2 + r^2 d\theta^2 +r^2\sin^2\theta \, d\phi^2 \,.
\eea

The generator of the horizon \reef{horgen}, in terms of the coordinates
of asymptotic observers
at rest, is
\be
  \xi=\gamma \left( \frac{\partial}{\partial \bar{t}}
  -v_H \frac{\partial}{\partial \bar{z}} \right) \, .
\ee
Hence the horizon is moving with velocity $v_H$ along $\bar{z}$ with respect to
asymptotic infinity.
So $\gamma$ is actually the usual relativistic dilation factor, which
was the motivation for its definition. Note also the explicit relativistic
form of the coordinate transformation \reef{asymptcoord}.

As can be seen from the metric \reef{asym}, the requirement $v_H<1$ is necessary
in order to avoid closed causal curves near infinity.
Observe also from \reef{asym2} that the physical
radius $R_5$ of the five-dimensional Kaluza-Klein circle at infinity,
$\bar{z}\sim \bar{z}+2\pi R_5$, is not that of the base space, $R_k$, but is contracted
by a factor
\be
R_5=R_k\gamma^{-1}\, .
\ee

\subsection{Absence of Closed Timelike Curves} \label{s:noCTCs}
In the two previous sections we have found two conditions necessary to
ensure that the spacetime is causally well-behaved. The condition
\be \label{noCTCasymp}
  v_H=\frac{3q}{4 \, (Q_k +R)}<1
\ee
is needed in order to avoid closed causal curves near asymptotic infinity.
Physically this condition means that the horizon cannot rotate faster than
light relative to an observer at rest at infinity.

On the other hand, we have found that in order to avoid CTCs near the
horizon, the condition \reef{Qcondition} must be obeyed. It
is shown in appendix \ref{app:noCTCs} that \reef{Qcondition} and
\reef{noCTCasymp} are sufficient conditions to ensure that the spacetime
has no CTCs outside the horizon.

\subsection{Physical parameters} \label{s:5dcharges}

The physical parameters of the solution can be computed by taking either a
five-dimensional viewpoint, in terms of quantities that belong in the minimal
five-dimensional supergravity, or in terms of the four-dimensional theory that is
obtained via KK reduction. Both kinds of magnitudes are of course directly related
(the relation is given in section \ref{s:4dred} below), but the
five-dimensional theory is simpler and so we take this viewpoint in this section.

The spacetime has a translation-invariant direction $\bar{z}$ and hence
it can be assigned a $2\times2$ ADM stress-energy tensor $T_{ab}$,
$a,b=\bar{t},\bar{z}$. This is computed from the asymptotic metric
$g_{\mu\nu}=\eta_{\mu\nu}+h_{\mu\nu}$ (in
Cartesian coordinates) as
\be
T_{ab}=\frac{1}{16\pi G_5}\int d\Omega_{(2)}r^2
n^i\left[\eta_{ab}\left(\partial_i h^c_c +\partial_i h^j_j -\partial_j h^j_i\right)-
\partial_i h_{ab}\right)]
\ee
where $n^i$ is the radial normal vector and $a,b,c$ run over parallel
directions $\bar{t},\bar{z}$ while $i,j$ run over transverse directions.
The integration is over the transverse angular directions. Using this we obtain
the mass and momentum as the integrated energy and momentum
densities,\footnote{The ubiquitous presence of the factor $\pi/(4G_5)$
can be eliminated by judiciously setting to one the eleven-dimensional
Planck constant,
$\ell_{11}=(4G_5/\pi)^{1/3}=1$.}
\be\label{mass}
M=\int d\bar{z}\, T_{\bar{t}\bar{t}}=\frac{\pi}{4G_5}\frac{\gamma}{N_k}\left(
3Q+4 Q_k^2 (1-2v_H^2)+8v_H^2 Q_k R
\right)\,,
\ee
\be\label{mom}
P=\int d\bar{z}\, T_{\bar{t}\bar{z}}=\frac{\pi}{4G_5}\frac{\gamma v_H}{N_k}\left(
3Q-4v_H^2 Q_k^2+8R Q_k
\right)\,.
\ee

The electric charge is defined as
\be\label{elcharge}
\mathbf{Q}=\frac{1}{16\pi G_5}\int_S \left( \frac{2}{\sqrt{3}} \star
\calf - \frac{4}{3} \cala \wedge \calf \right),
\ee
where $S$ denotes the $S^3/\mathbb{Z}_{N_k}$ at infinity and we have
chosen to absorb the awkward factor $2/\sqrt{3}$ to eliminate it
from the charge. The presence of the Chern-Simons term is required
for the charge to be conserved.\footnote{This term vanishes in an
asymptotically flat context but it is non-vanishing here.} This term
can be computed in any gauge regular in a neighbourhood of infinity.
It is gauge-invariant because the pull-backs to $S$ of any two gauge
potentials must differ by an exact form on $S$ (since $S$ has
vanishing first cohomology). One must be careful to work in the
frame that is at rest at infinity as defined above. We find
\be
\mathbf{Q}=\frac{\pi}{4G_5}\frac{Q}{N_k}\,.
\ee
We define the Kaluza-Klein monopole charge in a manner motivated by its
four-dimensional interpretation,
\be\label{kkmcharge}
\mathbf{Q}_{k}=\frac{2\pi R_5}{4G_5}\frac{N_k R_5}{2}=\frac{\pi}{4G_5}
\frac{4Q_k^2}{N_k\gamma^2}\,.
\ee
Then we have
\be\label{bps}
 E_{\xi} \equiv \gamma \left( M - v_H P \right) = 3 \mathbf{Q} + \mathbf{Q}_k,
\ee
which is presumably the BPS relation for this system. Note that the LHS
is the energy defined with respect to the supersymmetric (co-moving)
Killing field $\xi$ \reef{horgen}.

We can compute the dipole charge using the 5D formula
\be
\mathbf{D}=\frac{1}{16\pi
G_5}\int_{S^2}\frac{2}{\sqrt{3}}\calf=\frac{q}{8G_5}\,,
\ee
where the $S^2$ links the ring only once (one such surface is obtained
at constant $t,y,\hat\psi$, in the coordinates introduced in appendix
\ref{app:xy}). Since the topology of the base space is $\CR^4$, the
black ring does not wrap any non-trivial cycle and it can be
stripped of this charge. However, and in contrast to the situation
for the asymptotically flat 5D rings, in the Taub-NUT background
this dipole charge {\em is} conserved.\footnote{Hence the term
`dipole' is a bit of a misnomer, which we keep only for its
connection with the solutions in
\cite{RE,EEMR1}. In fact, the field created by this charge in four
dimensions behaves at large distances like a monopole field.} As
explained in \cite{GHM}, when the source of this charge is unwound a
current appears that takes the charge towards the nut. Taub-NUT
space possesses a harmonic anti-self-dual two-form \cite{GR} that
gives rise to a zero-mode of the gauge field. This gets excited in
the process of unwinding and as a result the nut acquires the
charge.\footnote{It must be noted, though, that for the bound state
that the supersymmetric black ring represents, the M5-branes cannot
be unwound and separated from the M2-branes without breaking
supersymmetry.}

Finally, in five dimensions the rotation along $\phi$ gives rise to an
angular momentum
density $\calj_\phi$ which is obtained from
\be
-\frac{g_{\bar{t}\phi}}{g_{\bar{t}\bar{t}}}=
 \frac{2G_5\calj_\phi \sin^2\theta}{r}+O(r^{-2})
\ee
so the angular momentum is
\bea\label{jphi}
J_\phi&=&\int d\bar{z}\, \calj_\phi=\frac{\pi}{G_5} \gamma v_H R R_5
Q_k\nn
&=& 2 v_H\gamma^2 R\mathbf{Q}_k\,.
\eea
The last expression can be used to eliminate the parameter $R$ (which
does not correspond to any invariantly defined quantity)
and obtain the relation
\be\label{dirac0}
\frac{1}{2}\mathbf{Q}_k
\left(P -3 \gamma v_H\mathbf{Q}+\gamma^3 v_H^3 \mathbf{Q}_k\right)
=\frac{\pi R_5}{4G_5}J_\phi\,.
\ee
In the limit where the ring and the nut are very far apart, $R$ being much
larger than all charges, we have $v_H\to 0$ and $\gamma\to 1$ so
this equation becomes,
\be\label{diracQP}
\frac{1}{2}\mathbf{Q}_k P
=\frac{\pi R_5}{4G_5}J_\phi\,.
\ee
$P$ and $\mathbf{Q}_k$ are actually quantized. The quantization of the KK
magnetic charge has already been given in \reef{kkmcharge},
\be
N_k=\frac{4G_5}{\pi}\frac{\mathbf{Q}_k}{R_5^2}
\ee
and the momentum is quantized as
\be
N_p=R_5 P\,.
\ee
Then \reef{diracQP} is
\be\label{dirac1}
\frac{N_p N_k}{2}=J_\phi\,.
\ee
This is the Dirac formula for the Poynting angular momentum created
by the electric and magnetic KK charges. For finite values of $R$,
however, there are additional terms in \reef{dirac0} that modify the
`effective electric charge' and which deserve to be better
understood.

\setcounter{equation}{0}
\section{Limits}\label{s:limits}

We study here two limits of the black ring: first the $R\to 0$ limit
where the solution describes a spherical black hole at the nut, secondly
the limit $R\to \infty$ yielding a compactified black string.

\subsection{Black hole at the nut} \label{s:bhatnut}
In the limit $R \to 0$ the black ring collapses to a spherical black
hole at the nut. In this case,
\be\label{Hatnut}
H=1+\frac{Q-\frac{3}{2}q^2}{4Q_k r}+\frac{q^2}{4H_k r^2}\,,
\ee
\be\label{omatnut}
\omega_0=v_H\left(1 - \frac{Q_k}{H_k r}
-\frac{Q-\frac{3}{2}q^2}{4r^2 H_k}-\frac{q^2
Q_k}{6r^3 H_k^2}\right)
\,,
\ee
and
\be
\ti{\omega}_\phi=0\,.
\ee
The latter implies that $J_\phi=0$ and that $\omega$ is aligned with
the TN fiber --- hence the reduction to four-dimensions results in a
static solution. A regular horizon is located at $r=0$ and there the
solution is similar to the BMPV black hole. The latter is actually recovered in
the limit $Q_k\to \infty$, $r\to 0$, $z\to\infty$ with new finite
coordinates $\rho^2=4Q_k r$ and $\psi=z/Q_k$.

For later purposes in this paper it will suffice to discuss in this
section the simpler case of a static black hole at the nut. This is
obtained by taking $R=0$ and $q=0$. Note that $\omega=\beta=0$
so this is a static electrically charged solution in the Taub-NUT
background. The metric is\footnote{The same solution is obtained
setting $J=0$ in the solution in \cite{GSY}.}
\be
ds^2=-H^{-2}
dt^2+\frac{H}{H_k}\left(dz+Q_k\cos\theta d\phi\right)^2
+HH_k\left(dr^2+r^2d\Omega^2_{(2)}\right)
\ee
with $H=1+\bar{Q}/r$ and $\bar{Q}\equiv Q/(4Q_k)$.
This is one of the possible uplifts to five-dimensions of the familiar
supersymmetric four-charge black hole (with three of the charges set
equal to $Q$), and in particular when $\bar{Q}=Q_k$ it reduces in 4D to the
extremal Reissner-Nordstr\"om solution. The horizon at $r=0$ is regular,
and the KK circle does not shrink to zero at $r=0$ but remains finite.
Hence $r=0$ will also remain a regular horizon in 4D, and this will be
used later to hide the singularity of the KK monopole in four
dimensions.

Near the horizon the solution becomes locally AdS$_2$ of radius
$\sqrt{Q}/2$ times a lens space $S^3/\mathbb{Z}_{N_k}$ of radius
$\sqrt{Q}$. The spatial cross-section of the horizon has topology
$S^3/\mathbb{Z}_{N_k}$ and the area is
\bea \label{AHnut5d}
  A_H = \frac{2\pi^2 Q^{3/2}}{N_k} \, .
\eea
This is not the same as the $q, R\to 0$ limit of the black ring horizon
area given in \eqn{ringarea}. The same phenomenon was discussed in
\cite{EEMR1,EEMR2}, where it was argued to be due to the different
topology of the horizons, and is analogous to the discontinuity in the
horizon area in the merger of two supersymmetric Reissner-Nordstr\"om black
holes. Nevertheless the physical charges measured at infinity are
correctly obtained by setting $R=q=0$ in the expressions in section
\ref{s:5dcharges}.

\subsection{Compactified black string}

In the limit $R\to\infty$ where the nut is moved infinitely far away
from the black ring we recover a compactified version of the
supersymmetric black string found in \cite{bena}.

To take this limit it is convenient to
choose coordinates in the base space that are centered at the ring
instead of at the nut. Denoting these
coordinates with a bar, they are
\be
\bar{r}=\Delta\,,\qquad
\sin\bar{\theta}=\frac{r}{\Delta}\sin\theta\,.
\ee
Then take $R\to\infty$ keeping $\bar{r}$ and $\bar{\theta}$ finite. In
this limit $H_k\to 1$ and
\be
H\to 1+\frac{\bar{Q}}{\bar{r}}+\frac{q^2}{4\bar{r}^2}\,,\qquad
\omega_0\to -\frac{3q}{2\bar{r}}-\frac{3q\bar{Q}}{4\bar{r}^2}-\frac{q^3}{8\bar{r}^3}
\ee
where, again, $\bar{Q}\equiv Q/(4Q_k)$. Since $\cos\theta\to-1$, and consequently
$(dz+Q_k\cos\theta d\phi)\to (dz-Q_kd\phi)$, it is convenient to
gauge-transform the string away, $z\to z+Q_k\phi$. Then one reproduces
the solution in \cite{bena}, compactified on a circle with period
$\Delta z=4\pi Q_k/N_k$.

Nevertheless, the effects of the topology of Taub-NUT on the ring, even
at arbitrarily large but finite $R$, make it significantly different
than the compactified black string in which the nut is absent. The ring
in Taub-NUT can always be unwound (even if it remains arbitrarily far
from the nut \cite{GHM}) whereas the black string of \cite{bena} cannot.
This is because the limit above changes the topology of the base space
from $\CR^4$ to $\CR^3 \times S^1$, and the string wraps a
non-contractible cycle. Another difference is that the black string has
$J_\phi=0$ whereas for the ring in TN space the angular momentum
\reef{dirac1} does not approach zero even at arbitrarily large values of
$R$. Again, the reason is topological: Dirac's result for the angular
momentum \reef{dirac1} is a topological linking number which does not
depend on the distance between the electric and magnetic poles.

\setcounter{equation}{0}
\section{Supersymmetric four-dimensional black holes} \label{s:4dred}

In the background of Taub-NUT space, the circle direction along the ring
becomes a compact dimension and the solution naturally reduces a la KK
to a four-dimensional black hole with angular momentum in the direction
$\phi$. It would be incorrect, though, to think that this angular
momentum results directly from the rotation of the $S^2$ in the black
ring. The $\phi$ direction of the four-dimensional black hole is {\em not} the
same as the azimuthal angle $\hat\phi$ for the $S^2$ of the black ring
(introduced in appendix \ref{app:xy}), but it is instead
$\phi=\hat\phi-\hat\psi$, where $\hat\psi$ is the angular direction
along the $S^1$ of the ring. So, if we want to interpret the rotation of
the four-dimensional black hole in terms of the rotation of the
five-dimensional black ring, the correct relation is
$J_\phi\leftrightarrow J_{\hat\phi}-J_{\hat\psi}$. On the other hand, since
$z=Q_k(\hat\psi+\hat\phi)$, the momentum $P$ actually corresponds to the
combination $J_{\hat\psi}+J_{\hat\phi}$.

In the reduction to four dimensions we get not only a black hole at $r=R$,
$\theta=\pi$, which carries, among
other charges, a KK electric charge from its motion in the $\bar{z}$
direction. We
also obtain a KK magnetic monopole at $r=0$ from the reduction of the
Taub-NUT geometry. Then, from the point of view of four-dimensional
physics, it is the presence of both electric and magnetic KK charges
that gives rise to rotation in four dimensions.

The KK monopole is a naked singularity from the four-dimensional point
of view. This,
however, is not an essential difficulty since we know that in five
dimensions the
curvature singularity disappears (leaving only a well-understood
${\mathbb Z}_{N_k}$ orbifold
singularity). Moreover, in four dimensions this singularity can be
hidden by placing a
black hole at the origin of the Taub-NUT space, as we will show in the
next section.
The basic physics of the solution, though, is already exhibited in the
solution with
a naked nut, which we analyze now.

\subsection{Reduction of the 5D action}
The nutty black rings are solutions of the equations of motion
obtained from the action of ${\cal N}=1$ $D=5$ minimal supergravity,
\bea
  S_\rom{5d} = \frac{1}{16 \pi G_5} \left[
  \int d^5x \sqrt{-g_{(5)}} \Big( R_{(5)} - \mc{F}^2 \Big)
  -\frac{8}{3 \sqrt{3}} \int \mc{F}\wedge\mc{F}\wedge\mc{A}
  \right] \, ,
\eea
where $\mc{F}=d\mc{A}$. Writing
\bea
  ds_\rom{5d}^2=e^{-2\Phi/\sqrt{3}}(dz+C_\mu dx^\mu)^2
  +e^{\Phi/\sqrt{3}} ds_\rom{4d}^2 \, ,
\eea
so the four-dimensional metric is in the Einstein frame,
and
$\mc{A} = A_\mu dx^\mu + \rho\, dz$, we find for the four-dimensional
action
\bea\label{action4d}
  \non
  S_\rom{4d} &=& \frac{1}{16 \pi G_4} \Bigg[
  \int d^4x \sqrt{-g} \left( R
  - \frac{1}{2} (\p\Phi)^2
  - 2 e^{2\Phi/\sqrt{3}} (\p\rho)^2
  - \frac{1}{4} e^{-\sqrt{3}\Phi} G^2
  - e^{-\Phi/\sqrt{3}} \ti{F}^2  \right)  \\
  &&\hspace{2cm}
  -\frac{8}{\sqrt{3}} \int \rho\, F\wedge F
  \Bigg] \, .
\eea
where
\bea
  G_4=G_5/(2\pi R_5) \, ,~~~~
  G=dC \, ,~~~~
  F=dA \, , ~~~\mbox{and}~~~
  \ti{F}=F+C\wedge d\rho \, .
\eea
The Chern-Simons term in $\ti{F}$ comes from the inverse
five-dimensional metric.

A consistent truncation of this theory is obtained by setting
\be
 \rho=0=\Phi\,,\qquad \star G = +\- \frac{2}{\sqrt{3}} F\,,
\ee
which reduces \reef{action4d} to the Einstein-Maxwell theory.

\subsection{The 4D solution}\label{s:4dsoln}

The five-dimensional metric can be written in the form
\bea \label{5dmetric}
  ds_\rom{5d}^2 = \left(\frac{U}{H H_k}\right)^2 \bigg[ dz - S\, dt
    + T\, d\phi \bigg]^2 + d\ti{s}_\rom{4d}^2
\eea
where
\bea \label{4dmetric}
  d\ti{s}_\rom{4d}^2 =
  - \frac{H H_k}{U^2} \Big[ dt+\ti{\omega}_\phi\,d\phi\Big]^2
  + H H_k \Big[ dr^2 + r^2\,d\theta^2 + r^2 \sin^2\theta d\phi^2
  \Big]\, ,
\eea
with
\bea\label{UST}
  U&=& \sqrt{H^3 H_k- \omega_0^2 H_k^2}\, ,\\
  S &=& \omega_0 H_k^2 U^{-2} \, , \\
  T &=& Q_k \cos\theta
      - \omega_0 \ti{\omega}_\phi H_k^2 U^{-2} \, .
\eea

We introduce now the asymptotic five-dimensional coordinate $\bar{z}$ and the
canonical time coordinate $\bar{t}$ defined in
\reef{asymptcoord}. Then
\bea
  ds_\rom{5d}^2 = \gamma^2 \left(\frac{U}{H H_k}\right)^2
   \bigg[ d\bar{z} + \left(v_H-\gamma^{-2}S\right)\, d\bar{t}
    + \gamma^{-1}T\, d\phi \bigg]^2
  + d\ti{s}_\rom{4d}^2  \, ,
\eea
hence
\be
  e^{-\Phi/\sqrt{3}}=\gamma\frac{U}{HH_k}\,,
\ee
the Einstein metric is
\be
ds_\rom{4d}^2=
- \frac{1}{\gamma U}\Big[
d\bar{t}+\gamma\ti{\omega}_\phi\,d\phi\Big]^2
+ \gamma U \Big[ dr^2 + r^2\,d\theta^2 +
  r^2 \sin^2\theta d\phi^2
  \Big]\, ,
\ee
and the gauge field $C$ has components
\bea
  C_{\bar{t}} =v_H-\gamma^{-2}S \, ,~~~~~
  C_\phi = \gamma^{-1}T \, .
\eea

The five-dimensional gauge potential $\mc{A}$ gives rise to the
four-dimensional pseudoscalar (axion)
$\rho = \mc{A}_{\bar{z}}$ given by
\bea
  \rho = \frac{\sqrt{3}}{2} \, \gamma
  \left( H^{-1} \omega_0 - \beta_0 \right)
\eea
and the gauge potential $A$, taking into account the shift \reef{asymptcoord},
\bea
  A_{\bar{t}} &=&  \frac{\sqrt{3}}{2\gamma}
  \Big(H^{-1}+v_H \gamma^2
  \left( H^{-1} \omega_0 - \beta_0 \right) \Big)\, ,\nn
  A_\phi &=& \frac{\sqrt{3}}{2}
  \left( H^{-1} (\omega_0 \, Q_k \cos\t +\ti{\omega}_\phi)
  -(\beta_0 \, Q_k \cos\t +\ti{\beta}_\phi)
  \right) \, .
\eea

The mass and angular momentum (measured in the Einstein metric) are
the same as $M$ in \reef{mass} and $J_\phi$ in \reef{jphi}.
The KK electric and magnetic charges correspond to $P$ and
$\mathbf{Q}_k$,
\be\label{kkcharges}
\mathbf{Q}_e=\frac{1}{16\pi G_4}\int \star\; e^{-\sqrt{3}\Phi}G=P\,,\qquad
\mathbf{Q}_m=\frac{1}{16\pi G_4}\int G=\mathbf{Q}_k\,.
\ee
The charge $\mathbf{Q}$ in \reef{elcharge} is the same as
\be
\mathbf{Q}=\frac{1}{16\pi G_4}\int_{S^2}\left(\star\;\frac{2}{\sqrt{3}} e^{-
\Phi/\sqrt{3}}\ti{F}+\frac{4}{3}(\rho F-d\rho\wedge A)\right)\,.
\ee
The magnetic dual of this charge is
\be\label{magP}
\mathbf{P} = \frac{1}{16 \pi G_4} \int_{S^2} \frac{2}{\sqrt{3}} \left(
\tilde{F} - \rho G \right) = \frac{q}{8G_4} = 2\pi R_5 \mathbf{D}\,,
\ee
which makes explicit that the dipole charge is, in this system, a
conserved charge as explained in section \ref{s:5dcharges}.

\subsubsection*{Limit $R\to 0$}

We have argued in section \ref{s:bhatnut} that in the limit $R\to 0$ we
obtain a black hole at the nut which, for $q>0$, is non-static
in five-dimensions but static in four.

A related, but different, black hole was considered in \cite{GSY}. In
five dimensions it also describes a BMPV-like black hole at the nut in a
TN background, and it is also obtained using the methods of
\cite{harveyetal}. Using our notation, it corresponds to taking, for
$K,L,M$ in \reef{LKM}, the functions
\be
K=0\,,\quad L=1+\frac{Q}{4Q_k r}\,,\qquad M=v_H H_k
\ee
with $H_k=1+Q_k/r$. When reduced to four dimensions, this is a static
black hole with non-zero charges $\mathbf{Q}_e$, $\mathbf{Q}_m$ and
$\mathbf{Q}$, but, since $\ti\beta=0$, it has zero magnetic charge
$\mathbf{P}$. This charge distinguishes it from our solution.

\setcounter{equation}{0}
\section{Rotating BPS two-black hole solutions} \label{s:coverNUT}

The goal of this section is to construct a solution with two black
holes: in five dimensions we place a black hole at the nut and a black
ring away from the nut. In four dimensions this gives a
supersymmetric configuration with two black holes and a net angular
momentum. This solution is regular with no singularities
outside the horizons.

More generally, one can construct multi-concentric supersymmetric
black ring configurations in Taub-NUT, just as it has been done
for asymptotically flat black rings \cite{GG1,GG2}. When all the rings
wrap the TN fiber, the reduction to 4D describes a sequence of
supersymmetric black holes along the axis of rotation to one side of the
nut.

\subsection{Solution}
We shall construct the simplest example of a black hole at the nut and a
black ring away from the nut.
Take
\bea
  K = -\frac{q}{2\Delta}  \, ,~~~~
  L = 1 + \frac{Q_\mathrm{r}-2qv_H Q_k}{4Q_k\Delta}
  + \frac{Q_\mathrm{h}}{4Q_k r} \, ,~~~~
  M = v_H \left( 1 -
  \frac{R+\frac{Q_\mathrm{h}}{4Q_k}}{1-\frac{Q_\mathrm{h}}{4Q_k^2}}\frac{1}{\Delta}
  \right) \, ,
\eea
where $Q_\mathrm{r}$ and $Q_\mathrm{h}$ are constants and
\bea \label{vH2bh}
  v_H = \frac{3q}{4\, (Q_k+R)} \left( 1-\frac{Q_\mathrm{h}}{4Q_k^2}\right) \, .
\eea
The constants in $M$ are fixed by
requiring that $\ti{\omega}_\phi$ vanishes as $r\to \infty$ (necessary
for asymptotic flatness in four dimensions) and that
$\ti{\omega}_\phi(\theta=0,\pi)=0$ (to avoid Dirac-Misner strings in
five dimensions).

The parameters $Q_\mathrm{r}$ and $Q_\mathrm{h}$
play the roles of the charge parameters for the ring and the hole,
respectively. When $Q_\mathrm{h}=0$ we recover the solution for a black ring in
TN of section \ref{s:soln}, while $Q_\mathrm{r}=0=q$ yields the static black hole
at the nut of section \ref{s:bhatnut}.

With the above choice, we have
\bea
  H = 1 + \frac{Q_\mathrm{h}}{4Q_k\,r}
  +\frac{Q_\mathrm{r}-2qv_H Q_k}{4Q_k\Delta}
  + \frac{q^2}{4 H_k \Delta^2} \, ,
\eea
\be
\omega_0=v_H \left( 1 -
  \frac{R+\frac{Q_\mathrm{h}}{4Q_k}}{1-\frac{Q_\mathrm{h}}{4Q_k^2}}\frac{1}{\Delta}
  \right)-
\frac{3q}{4H_k \Delta}\left(1 + \frac{Q_\mathrm{h}}{4Q_k\,r}+
\frac{Q- 2 q v_H Q_k}{4Q_k \Delta}\right)
-\frac{q^3}{8H_k^2\Delta^3}\, ,
\ee
and
\be
  \tilde{\omega}_\phi =
  \frac{2 v_H\,Q_k\,R\,r\,\sin^2\theta}
     {\Delta\,(r+R+\Delta)} \, .
\ee
Finally, $\beta$ takes the same form as in \reef{beta}. The solution is
then obtained by plugging these functions into \reef{metric},
\reef{potential} and \reef{beom}.

\subsection{Properties}

\subsubsection*{Asymptotics and physical charges}

It was shown in section \ref{s:5dasymp} that a coordinate
transformation is needed to bring the asymptotic metric to a frame at
rest. For the two-black hole solution here this is done just as in
section \ref{s:5dasymp}, but now with $v_H$ as given in
\reef{vH2bh}. The necessary condition for avoiding CTCs near
asymptotic infinity is $v_H<1$.

The dipole charge of the black ring is $\mathbf{D}=q/(8G_5)$.
For the conserved charge \reef{elcharge} we find
\bea
  \mathbf{Q}=\frac{\pi}{4 G_5 N_k}
  \left[ Q_\mathrm{h} + Q_\mathrm{r} \right] \, .
\eea
Note, however, that it is not rigorous to associate $Q_\mathrm{h}$
and $Q_\mathrm{r}$ with the charges of the black hole and the black
ring separately, since only the total charge $\mathbf{Q}$ is
measured at infinity. Presumably, one could use the method of
\cite{HR} to assign separate charges to the hole and the ring.

\subsubsection*{Near-horizon geometry}
The analysis of the near-horizon geometries goes through for both the
black hole at the nut and the black ring just as for the single-black
hole solutions studied in the previous sections.

The black hole at the nut has horizon
$S^3/\mathbb{Z}_{N_k}$ with radius
$\sqrt{Q_\mathrm{h}}$. The black ring has an $S^1 \times S^2$
horizon. The radius of the $S^2$ is $q/2$ and the $S^1$ has proper
length $\hat{L}_r$, where
\bea
  \hat{L}_r = \sqrt{3\left( \frac{(Q_\mathrm{r}-2qv_H Q_k)^2}{4 q^2} -
    \frac{Q_k(4RQ_k+Q_\mathrm{h})}{R+Q_k}\right)} \, .
\eea
The horizon areas are
\bea
  A_H^\rom{h} = \frac{2\pi^2 Q_\mathrm{h}^{3/2}}{N_k}\, ,~~~~~
  A_H^\rom{r} = \frac{2\pi^2 q^2 \hat{L}_r}{N_k}\, .
\eea

\subsubsection*{Absence of CTCs}
To avoid CTCs near asymptotic infinity, we must require $v_H<1$, \ie
\bea \label{nctc2bh1}
  3 q\,(4Q_k^2-Q_\mathrm{h}) < 16Q_k^2 \, (Q_k+R) \, .
\eea
Near the black hole horizons the absence of CTCs requires
\bea \label{nctc2bh2}
  Q_\mathrm{h} > 0\, ,~~~~~~~
   Q_\mathrm{r}-2qv_H Q_k  \ge 2q \,\sqrt{\frac{Q_k(4RQ_k+Q_\mathrm{h})}{R+Q_k}}  \, .
\eea
The three conditions in equations \reef{nctc2bh1} and \reef{nctc2bh2}
are necessary for absence of CTCs outside the horizons. We have not
proven that these conditions are sufficient. However, our proof that
there are no CTCs for $Q_\mathrm{h}=0$ actually
implies that the eigenvalues of the relevant $2\times 2$ matrix are strictly
positive provided the appropriate bounds are strictly enforced, \ie
provided these bounds are not saturated. By continuity, it follows that
those eigenvalues remain positive for small enough $Q_\mathrm{h}$, and hence there are
no CTCs for a small enough black hole at the nut.

\subsubsection*{Reduction to four dimensions}
The reduction of the black
hole + black ring solution yields two black holes in four dimensions,
one at $r=0$, the other at $r=R$.
The explicit form of the metric and other fields can be obtained using the
same formulas as in section \ref{s:4dsoln}. The horizons are smooth and
the nut singularity is hidden behind the horizon of one of the two black
holes. The area of the horizons are easily derived from the
five-dimensional areas given above.

The most exciting feature of this solution is that it has a net angular
momentum, $G_4 J_\phi = \frac{1}{2} Q_k R v_h \gamma$. This is an
explicit example of a supersymmetric four-dimensional multi-black hole
configuration with angular momentum and without naked singularities nor
any other pathologies.

\subsection{Generalizations}\label{s:general}
A general starting point for finding a solution with a black hole at
the nut and a black ring away from the nut is given by
\bea
  K = k_1 + \frac{k_2}{\Delta} + \frac{k_3}{r} \, ,~~~~
  L = l_1 + \frac{l_2}{\Delta} + \frac{l_3}{r} \, ,~~~~
  M = m_1 + \frac{m_2}{\Delta} + \frac{m_3}{r} \, .
\eea

Above we have studied for simplicity the case $k_1=m_3=0$, but one could
consider keeping these parameters. In particular, it is interesting that when
$k_1\neq 0$ one can obtain rings with vanishing horizon velocity $v_H$.
To see this, consider the generalization
of the solutions of section \ref{s:soln} with $k_3=l_3=m_3=0$ (\ie with
a naked nut) but $k_1\neq 0$.  A straightforward extension of
our analysis in section \ref{s:soln} shows that if we require
$\ti\omega_\phi$ not to have Dirac-Misner strings and to
vanish at $r\to\infty$, then the conditions
\be\label{ems}
m_2=-m_1 R
\ee
and
\be\label{kl}
k_1 l_2-k_2 l_1=\frac{2 m_1}{3}\left(R+Q_k\right)
\ee
must be imposed. The asymptotic behavior of $\omega_0$ implies
that the velocity of the horizon is
\be\label{vh}
v_H=m_1 +\frac{3}{2}k_1 l_1 +k_1^3\,.
\ee
Asymptotic flatness also demands that $\gamma U\to 1$ as $r\to\infty$ (with
$U$ defined in \reef{UST} and $\gamma$ in \reef{gamdef}) and
this translates into
\be\label{gamu}
l_1+k_1^2=1\,.
\ee
For generic $v_H$, these conditions leave three of the six parameters
$l_i,k_i,m_i$ arbitrary. The explicit form of $\ti\omega_\phi$ is
obtained by changing $v_H\to m_1$ in \reef{tilde}.

Near the horizon we find the same structure as in \reef{nhmetric} but
now
\be
\hat{L}^2=\frac{Q_k^2}{k_2^2}\left(3 l_2^2-8k_2 m_2\right)\,,
\ee
and the radius of the $S^2$ is $|k_2|$ instead of $q/2$. Absence of
causal pathologies requires $0\leq v_H <1$ and $\hat{L}^2>0$.

Our choice of $k_1=0$ in section \reef{s:soln} was made for simplicity
and served our purpose of building a black ring in TN space. But if
$k_1\neq 0$ we can have a non-static configuration (\ie $m_1,m_2\neq 0$)
with $v_H=0$. Imposition of \reef{ems}, \reef{kl}, \reef{vh},
\reef{gamu} leads to consistent solutions, but we have
not dwelt more on them.

Note furthermore that, besides the constraints
imposed by absence of CTCs, the parameters must be such that the mass be
positive. When $v_H=0$ one finds
\be
M=\frac{1}{4G_4}\left(3 l_2 +6k_1 k_2 +Q_k(1-3k_1^2)\right)\,
\ee
and it is
straightforward to
check the BPS relation $M=3\mathbf{Q}+\mathbf{Q}_k$. Presumably, \reef{bps}
extends to the generic case with $0<v_H<1$.

\setcounter{equation}{0}
\section{Discussion} \label{s:discussion}

We have constructed explicit solutions showing that five-dimensional
black rings are naturally related to supersymmetric configurations with
black holes and angular momentum in four dimensions. The solutions
exhibit a number of unusual features which we have only begun to
analyze. One of these is the fact that the horizon, despite being
supersymmetric, is generically moving at non-zero subluminal speed
$0<v_H<1$. This introduces a number of peculiar relations among
parameters, such as equation \reef{bps}, which we conjecture should be
understood as a BPS relation, and the relation \reef{dirac0}. The study
of the larger family of solutions described in section \ref{s:general},
which also allow for $v_H=0$, will probably shed light on the full
significance of these issues.

There is an interesting, simpler, related system, that exhibits
similar physics but without supersymmetry: a configuration
consisting of an electric and a magnetic KK black hole separated
apart---or, in string theory terms, a system of a D0-charged black
hole and a D6-charged black hole away from it. Again, a
non-vanishing electromagnetic angular momentum will be present. The
KK electric black hole would uplift to five dimensions as a neutral
black ring (like the original black ring in
\cite{ER}) which wraps the fiber of TN space, while the magnetic black
hole would uplift to a black hole sitting at the nut, or to a naked
nut in the extremal limit. It would be quite interesting to find
explicitly this solution of five-dimensional vacuum
gravity.\footnote{Such neutral solutions in five-dimensions (and
also generalizations of them to include dipole charges like in
\cite{RE}) should also be interesting to provide seeds for the
construction of near-extremal excitations of supertubes in Taub-NUT,
extending the work of \cite{previous}.} For this configuration,
however, all of the supersymmetry is broken, even when the extremal
limit is taken for the black holes.

An obvious generalization of the solutions in this paper is to unequal
charges, in general to a $U(1)^N$ theory, using a straightforward
application of the analysis in \cite{GG2}. These are the Taub-NUT
analogues of the three-charge supersymmetric black rings found in
\cite{EEMR1,BW,GG2}. Although the latter exhibit continuous
non-uniqueness (classically) of black rings, this is not the case for
the solutions in Taub-NUT. As we have explained, the dipole charges
responsible for the non-uniqueness in the 5D asymptotically flat
solutions become conserved charges in the Taub-NUT background. Upon
reduction to four dimensions they are conserved magnetic charges which,
together with the other charges, determine uniquely the black hole
solution.

It would be interesting to find a microscopic description, along the
lines of \cite{CGMS}, of the black ring in Taub-NUT in the limit where
the ring is very close to the nut. Then it effectively looks like a ring
in a $\mathbb{Z}_{N_k}$-orbifold of 5D space and its physical parameters
are simply rescaled by appropriate factors of $N_k$ (see \cite{GSY}).
However, the microscopic picture in \cite{CGMS} includes a zero-point
contribution to the oscillator number that apparently does not scale in
the correct manner.

Further insight into these solutions may also be obtained by studying
three-charge/three-dipole supertube probes \cite{EEMR2} in the Taub-NUT
background. Finally, other interesting extensions of this work include
the microscopic analysis of the system as a D1-D5-P configuration
following \cite{BK,BK3}, and the application to these solutions of the
attractor mechanism for black rings \cite{KL}.

 \medskip
\section*{Acknowledgments}
It is a pleasure to thank Gary Horowitz and Don Marolf for useful
discussions. We are particularly grateful to Frederik Denef for making
us aware of the work in \cite{denef}, which had been overlooked in the first
version of this paper. HE would like to thank the University of Michigan
and the University of Oregon for hospitality. HE was supported by NSF
grant PHY-0244764. RE was supported by CICYT FPA 2004-04582-C02-02,
DURSI 2001 SGR-00188 and European Comission FP6 Programme
MRTN-CT-2004-005104. HSR was supported in part by the National Science
Foundation under Grant No. PHY99-07949.
\noindent

\appendix

\section*{Appendices}

\setcounter{equation}{0}
\section{Taub-NUT black ring in $(x,y)$-coordinates} \label{app:xy}

The black ring in asymptotically flat 5D space was originally given in \cite{EEMR1}
using a convenient set of coordinates that foliate space with ring-shaped
surfaces. These can be
introduced as well for Taub-NUT rings, and in fact they are useful in order to
analyze the solution near the horizon. They also provide a simple way to recover the
asymptotically flat solution
of \cite{EEMR1}. These
$(x,y)$-coordinates are introduced through
\bea
  r = -R\frac{x+y}{x-y}\, , ~~~~~
  \cos\theta = -1+2\frac{1-x^2}{y^2-x^2}=1-2\frac{y^2-1}{y^2-x^2}\, ,
\eea
so that
\be
  \Delta = \frac{2 R}{x-y} \,.
\ee
A form of the solution that is more closely related to the one in
\cite{EEMR1} can be obtained as follows.
It is convenient to introduce
\be
  F_k=\frac{R}{Q_k}\frac{y+x}{y-x}\,H_k
  \; =\; 1+\frac{R}{Q_k}\frac{y+x}{y-x} \,.
\ee

The function $F_k$ encodes the effect of the Taub-NUT background on the
ring. Near the horizon, which lies at $y\to-\infty$, it will simply induce a
constant rescaling of magnitudes, but near asymptotic infinity, at
$y\to x\to-1$,
it diverges and hence has a crucial effect, namely the
compactification to finite radius of the $z$ circles.

The angular variables that parametrize the $S^1$ and the azimuth of the
$S^2$ of the ring in \cite{EEMR1} are not the same as $z$ and $\phi$.
We define new angular variables $\hat\psi$, $\hat\phi$, as
\be\label{hatpsiphi}
  \hat\phi=\frac{1}{2}\left(\frac{z}{Q_k}+\phi\right)\,,
  \qquad
  \hat\psi=\frac{1}{2}\left(\frac{z}{Q_k}-\phi\right) \, ,
\ee
which are identified as
\be
(\hat\psi,\hat\phi)\sim(\hat\psi,\hat\phi+2\pi)
\sim(\hat\psi+\frac{2\pi}{N_k}, \hat\phi+\frac{2\pi}{N_k}) \, .
\ee
We also define
\be\label{hatted}
\hat{R}\equiv 2 \sqrt{Q_k R}\,,\qquad
f_k\equiv\lim_{y\to-\infty} F_k=1+\frac{\hat{R}^2}{4Q_k^2}\,,\qquad
\hat{Q}\equiv Q-\frac{q^2}{2f_k^2}\left(1+\frac{3\hat{R}^2}{4Q_k^2}\right)\,.
\ee
$\hat{R}$ and $\hat{Q}$ correspond to
the radius scale and the charge, respectively, that were denoted by
$R$ and
$Q$ in \cite{EEMR1}. We find
\be
H=1+\frac{\hat{Q}-q^2/f_k^2}{2\hat{R}^2}(x-y)-\frac{q^2}{4\hat{R}^2}
\frac{x^2-y^2}{F_k}\,.
\ee
The base space takes on a symmetrical form
\bea\label{TNxy}
  ds^2(\rom{TN}) &=&
  \frac{\hat{R}^2}{(x-y)^2}\Biggl[F_k \left(\frac{dy^2}{y^2-
1}+\frac{dx^2}{1-x^2} \right)+
\frac{1}{F_k}\left((y^2-1)d\hat\psi^2+(1-x^2)d\hat\phi^2 \right)\nn
&&~~~~~~~~~+\left(F_k-F_k^{-1}\right) \frac{(y^2-1)(1-x^2)}{y^2-
x^2}\left(d\hat\psi-d\hat\phi\right)^2
\Biggr] \,.
\eea

In these coordinates the one-form $\omega$ has components
\be
\omega_{\hat\psi}=Q_k\omega_0(1-\cos\theta)-\ti\omega_\phi\,,\qquad
\omega_{\hat\phi}=Q_k\omega_0(1+\cos\theta)+\ti\omega_\phi\,.
\ee
These are
\be
  \omega_{\hat\psi}
  =\frac{3q}{2f_k}(1+y)
  -\frac{3q}{8 \hat{R}^2 F_k}(y^2-1)
\left[\hat{Q}-(q/f_k)^2-
\frac{q^2}{3F_k}(x+y)+
\frac{\hat{R}^4}{f_kQ_k^2}\frac{x}{(x-y)^2}\right]
\ee
and
\be
\omega_{\hat\phi}=-\frac{3q}{8\hat{R}^2 F_k}(1-x^2)
\left[\hat{Q}-(q/f_k)^2-
\frac{q^2}{3F_k}(x+y)+\frac{\hat{R}^4}{f_kQ_k^2}\frac{x}{(x-
y)^2}\right]\,.
\ee
Finally, the one-form $\beta$ is
\be
\beta =\frac{q}{2}\left[ (1+x)\left(1+(F_k^{-1}-1)\frac{1-x}{y-x}\right)d\hat{\phi}
+(1+y)\left(1+(F_k^{-1}-1)\frac{y-1}{y-x}\right)d\hat{\psi} \right]
\ee
Here an appropriate gauge choice has been made so that $\beta_{\hat\phi}$ vanishes
at $x=-1$ and $\beta_{\hat\psi}$ at $y=-1$.

Consider now the limit $Q_k\to\infty$ with fixed $\hat{R}$, $\hat{Q}$,
$q$, $x$ and $y$, \ie $F_k, f_k\to 1$. Then the metric \reef{TNxy}
becomes the $\bbr{4}$ base space in the form used for black rings in
asymptotically flat spacetime (with $N_k=1$ and identifying
$\hat{R},\hat\psi,\hat\phi$ with the corresponding unhatted variables in
\cite{EEMR1}). Indeed, it is straightforward to check that the complete
solution \reef{metric} and \reef{potential} reproduces in this limit the
supersymmetric black ring in \cite{EEMR1}.

\setcounter{equation}{0}
\section{Near-horizon geometry}\label{app:horizon}

Using the solution in $(x,y)$-coordinates as introduced in the previous appendix,
we study the near-horizon geometry $y\to -\infty$.

Near the horizon, define $\bar{r}=-\hat{R}/y$ and $\cos\bar\theta=x$.
Then the divergences near the horizon
are removed by defining new coordinates
\be
  dt=dv - B(\bar{r})\,d\bar{r},\qquad
  d\hat{\phi}=d\hat{\phi}'-C(\bar{r})\,d\bar{r}\,,\qquad
  d\hat{\psi}=d\hat{\psi}'-C(\bar{r})\,d\bar{r}\,,
\ee
with
\bea
  B(\bar{r}) = B_0+\frac{B_1}{\bar{r}}+\frac{B_2}{\bar{r}^2}\, ,~~~~~
  C(\bar{r}) = C_0+\frac{C_1}{\bar{r}}
\eea
and $B_i$ and $C_i$ are constants to be determined.
For the gauge field, we find
\bea
  \non
  \mc{A} &=&  \frac{2\sqrt{3} f_k}{q^2} \bar{r}^2 \Big( 1 + O(\bar{r}) \Big)dv
  + \frac{\sqrt{3}q}{4}
  \left( \cos\bar\theta + a_0 + O(\bar{r})\right)d\hat{\psi}'\\[2mm]
  && \hspace{1cm}
  - \frac{\sqrt{3}q}{4}
  \Big(1+\cos\bar\theta + O(\bar{r})\Big)d\hat{\phi}'
  + \left(\frac{b_1}{r} + b_0+ O(\bar{r})\right)d\bar{r} \, ,
\eea
where $a_0$, $b_0$, and $b_1$ are
constants. The $1/\bar{r}$-divergence in $\mc{A}_{\bar{r}}$ can be removed by a gauge
transformation.

The $1/\bar{r}$ and $1/\bar{r}^2$ divergences in $g_{\hat{\psi}'\bar{r}}$ and
$g_{\bar{r}\bar{r}}$
are cancelled by choosing $B_2=q^2 \hat{L}/(4\hat{R})$ and $C_1=-q/(2\hat{L})$,
where
\be\label{ELL2}
\hat{L}=\sqrt{3\left(\frac{(\hat{Q}-(q/f_k)^2)^2}{4q^2}-\hat{R}^2 f_k^{-1}\right)}\,.
\ee
This leaves a
$1/\bar{r}$-divergence in $g_{\bar{r}\bar{r}}$ which can been avoided by setting
\bea
  B_1 = \frac{\hat{Q}+2(q/f_k)^2}{4 \hat{L}}
        + f_k \frac{\hat{L}\,\big(\hat{Q}-(q/f_k)^2\big)}{3\hat{R}^2}
        + \frac{3 q^2 \hat{R}^2}{16 Q_k \hat{L} f_k^2} \, .
\eea
With the above choices, $g_{\bar{r}\bar{r}}$ is a linear
function of $x$ at $\bar{r}=0$. The constants $C_0$ and $B_0$ can be chosen such
that this linear function vanishes and
we have $g_{\bar{r}\bar{r}}\to O(\bar{r})$ when $\bar{r}\to 0$.

We can now write the near-horizon geometry
\bea
  \non
  ds_\rom{nh5d}^2 &=&
  - \frac{16 f_k^2 \bar{r}^4 }{q^4}dv^2
  + 2\frac{\hat{R}}{\hat{L}} dv d\bar{r}
  + \frac{4\bar{r}^3 \sin^2\bar{\theta}}{q \hat{R}}d\hat{\phi}'dv \\[2mm]
  &&
  +\frac{4 \hat{R} \bar{r}}{q} d\hat{\psi}'dv
  + \frac{3 q \bar{r} \sin^2\bar{\theta}}{\hat{L} f_k} d\bar{r}
  d\hat{\phi}' \\[2mm]
  \non
  &&
  +2\left(
   \frac{q \hat{L}}{2 \hat{R}} \cos\bar{\theta}
   + \frac{3 q \hat{R}}{2 \hat{L} f_k}
   +\frac{\hat{Q}-(q/f_k)^2}{6 \hat{R} \hat{L} f_k}
     \Big[ 3\hat{R}^2 (1+f_k) - 4f_k (2-f_k)\hat{L}^2 \Big]
   \right) d\hat{\psi}' d\bar{r}\\[2mm]
  \non
  &&
  + \hat{L}^2 d\hat{\psi}'^2
  + \frac{q^2}{4} \left[ d\bar{\theta}^2
  + \sin^2\bar{\theta} (d\hat{\psi}'-d\hat{\phi}')^2\right]
  + \dots
\eea
where ``$\dots$'' denotes subleading terms $O(\bar{r})$, including the
leading $O(\bar{r})$-terms in $g_{\bar{r}\bar{r}}$. We note that in
the $Q_k\to \infty$ limit we recover the near-horizon geometry of the
asymptotically flat black rings of \cite{EEMR1}.

Going back to $x=\cos\bar\theta$, the determinant of this metric is
$-q^4\hat{R}^2/16$ at $\bar{r}\to 0$, so both the metric and the inverse metric
are analytic at $\bar{r}=0$. The Killing vector $\partial_v$ is null at
$\bar{r}=0$ and is normal to the surface defined by $\bar{r}=0$. So the $\bar{r}=0$
null hypersurface is indeed the Killing event horizon for the black ring.

In the near-horizon limit we take $\epsilon\to 0$ while rescaling
$\bar{r}$ and $v$ as $\bar{r}=\ti{r} \hat{L} \epsilon/\hat{R}$ and
$v=\ti{v}/\epsilon$. The
metric in this limit is
\bea
  ds_\rom{nh}^2 = 2d\ti{v} d\ti{r}
  + \frac{4\hat{L}\ti{r}}{q} d\ti{v} d\hat{\psi}'
  + \hat{L}^2 d\hat{\psi}'^2
  + \frac{q^2}{4} \left[ d\bar\theta^2
  + \sin^2\bar\theta d\chi^2\right] \, .
\eea
We have denoted $\chi=\hat{\psi}'-\hat{\phi}'=\hat{\psi}-\hat{\phi}$.
The near-horizon limit of the metric describes the product of locally
AdS$_3$ of
radius $q$ with a two-sphere of radius $q/2$. This is the
same as was obtained for the ring in asymptotically flat 5D space in
\cite{EEMR1}, up to the substitution of $\hat{L}$ for $L$. In the limit
$Q_k\to\infty$ the results of \cite{EEMR1} are recovered.

\setcounter{equation}{0}
\section{Absence of Closed Timelike Curves}\label{app:noCTCs}

A sufficient condition for the metric \eqn{metric} with Taub-NUT base space to be free of
closed timelike curves is that the two-dimensional $z$-$\phi$ metric
\be
ds^2 = H^{-2} [ A\, (dz + Q_k \cos\theta \, d\phi)^2 + B \, d\phi^2
-2C \, (dz + Q_k \cos\theta \, d\phi) \, d\phi ]
\label{2Dmetric}
\ee
be positive-definite, where
\bea
A &=& H^3 \, H_k^{-1} - \omega_0^2 \,,
\nn B &=& H^3 \, H_k \, r^2 \sin^2 \theta - \tilde{\omega}_\phi^2
\,,
\nn C &=& \omega_0 \, \tilde{\omega}_\phi \,.
\eea
$H$ and $\omega_0$ are specified in equations \eqn{LKM} in terms of
$K, L$ and $M$, which, together with $\ti\omega_\phi$, are given in
equations \eqn{tnKLM}, \eqn{tilde}.

The metric \eqn{2Dmetric} is positive-definite if and only if the matrix
\be
\left(%
\begin{array}{cc}
  A & -C \\
  -C & B \\
\end{array}%
\right)
\ee
is positive-definite. This in turn is equivalent to the two
conditions
\be
A > 0 \sac A B -C^2 = H^3 H_k^{-1} (A\, H_k^2 \, r^2 \, \sin^2\theta
- \tilde{\omega}_\phi^2) >0 \,,
\ee
so it is enough to show that
\be
D \equiv A\, H_k^2 \, r^2 \, \sin^2\theta - \tilde{\omega}_\phi^2 > 0 \,.
\ee
In the limit $r \ra \infty$ we find
\be
D = r^2 \sin^2 \theta \,
\left(1 - v_H^2 \right) + {\cal O} (r) \,,
\ee
so as found in section \ref{s:5dasymp} a necessary condition to avoid
CTCs asymptotically is that $v_H<1$.

The near-horizon condition for avoiding CTCs is $\hat{L}^2>0$, \ie
from \reef{ELL2}
\be\label{near}
\hat{Q}- \tilde{q}^2 > 4 \, \tilde{q} \, \sqrt{f_k Q_k R}  \,.
\ee
where we have defined $\qt = q/f_k$ for convenience in this appendix,
and we are using the quantities $\hat{Q}$ and $f_k$ which were introduced
in \reef{hatted}.

We now show that these two necessary conditions are also
sufficient to ensure absence of CTCs.
We begin by noting that $D$ is a monotonically increasing function
of $\hat{Q}$. Indeed, since $\hat{Q}$ only enters through $L$, and we have
$dL/d\hat{Q} >0$, we just need to see that $dA/dL >0$. A direct
calculation yields
\be
A = H_k^{-2} \left( \frac{3 K^2 L^2}{4} -2 K^3 M \right)
+ H_k^{-1} \left( L^3 - 3K L M \right) - M^2 \,,
\ee
and hence
\be
\frac{d A}{d L} = \frac{3}{2} H_k^{-2} K^2 L
+ 3 H_k^{-1} \left( L^2 - K M \right) \,,
\ee
where
\be
L^2 - K M = 1 + \frac{3 \,f_k \,\tilde{q}^2+ 4(\hat{Q}-\tilde{q}^2)}{8\,Q_k\, \Delta}
+ \frac{(\hat{Q}-\tilde{q}^2)^2 - 6 \,\tilde{q}^2 f_k \,R\, Q_k}
{16\, Q_k^2\,\Delta^2} \,.
\ee
Both summands in $dA/dL$ are positive provided the bound \eqn{near}
holds.
Thus we just need to show that $D>0$ when
$\hat{Q}-\qt^2 = 4 \tilde{q} \sqrt{f_k Q_k R}$.

In this case one finds
\be
16 \, Q_k^4 \, \Delta^3 \, D = D_0
+ (1-x^2)\sum_{i=1}^{8} D_i \,,
\ee
where
\bea
D_0 &=& 12 \, \qt^2 \, Q_k^4 \, (1+x) \, \Delta \left[
3 \, \Delta \,(r+R-\Delta) +(1-x)\, r \,(r+R) \right] \,,
\nn D_1 &=& 3 \, r^2 \qt^4 (R+Q_k)^3 \,,
\nn D_2 &=& 48 \,r \, Q_k^3 \, \qt \,
\Delta^2 \, (Q_k + r)  \, \sqrt{f_k Q_k R} \,,
\nn D_3 &=& 2 \, \qt^3\, R\, Q_k\, r^2 \, (9\Delta +2R) \, \sqrt{f_k Q_k R} \,,
\nn D_4 &=&  2 \, \qt^3 \, Q_k^3 \, r
\, (9\Delta +3r - R) \, \sqrt{f_k Q_k R} \,,
\nn D_5 &=& 2 \, \qt^3 \, Q_k^2  \, r \,
\left[9 (r+R) \Delta + 5R\,r -R^2 \right]\, \sqrt{f_k Q_k R} \,,
\nn D_6 &=& Q_k^2\, r (r+Q_k) \, \Delta^3 \, ( 16 \, Q_k^2 - 9\qt^2) \,,
\nn D_7 &=& 12 \, \qt^2 Q_k^2\,R\, r^2 \, \Delta
\left( 2R +3 \Delta\right) \,,
\nn D_8 &=& 3 \, \qt^2 \, Q_k^3 \, r \, \Delta \,\left[
4R^2+12R\,r + 6 \Delta \,(3R+r) - 3 \Delta^2\right] \,.
\eea
Each $D_i\ge 0$. In particular $D_6\ge 0$ by virtue of the bound
$v_H<1$. $D_4$ and $D_5$ are seen to be positive by considering the
cases $r<R$ and $r>R$ separately.

We conclude that $D\ge 0$ whenever the two conditions \reef{noCTCasymp}
and \reef{Qcondition} hold. Thus, within this range of parameters, there
are no CTCs outside the horizon.


\end{document}